\documentclass[prb,superscriptaddress,twocolumn]{revtex4-1}
\usepackage{bbm}
\usepackage{graphicx}
\usepackage{dcolumn}
\usepackage{bm}
\usepackage{subfigure}
\usepackage{amsmath}
\usepackage{hyperref}
\hypersetup{
     colorlinks = true,
     citecolor  = magenta,  
     linkcolor  = magenta 
}

\graphicspath{{figs/}}

\usepackage{attachfile}

\begin{document}
\title{Realization of a Hopf insulator in circuit systems}

\author{Zhu Wang}
\thanks{Z. Wang and X.-T. Zeng contributed equally to this work.}
\affiliation{Wuhan Institute of Quantum Technology, Wuhan 430206, China}
\affiliation{School of Physics and Technology, Wuhan University, Wuhan 430072, China}

\author{Xu-Tao Zeng}
\thanks{Z. Wang and X.-T. Zeng contributed equally to this work.}
\affiliation{School of Physics, Beihang University, Beijing 100191, China}
\affiliation{School of Physics and Technology, Wuhan University, Wuhan 430072, China}

\author{Yuanchuan Biao}
\affiliation{Wuhan Institute of Quantum Technology, Wuhan 430206, China}
\affiliation{School of Physics and Technology, Wuhan University, Wuhan 430072, China}

\author{Zhongbo Yan}
\email{yanzhb5@mail.sysu.edu.cn}
\affiliation{Guangdong Provincial Key Laboratory of Magnetoelectric Physics and Devices, School of Physics, Sun Yat-sen University, Guangzhou 510275, China}

\author{Rui Yu}
\email{yurui@whu.edu.cn}
\affiliation{Wuhan Institute of Quantum Technology, Wuhan 430206, China}
\affiliation{School of Physics and Technology, Wuhan University, Wuhan 430072, China}

\date{\today}

\begin{abstract}
Three-dimensional (3D) two-band Hopf insulators are a paradigmatic example of topological phases beyond the topological classifications based on powerful methods like $K$-theory and symmetry indicators.
Since this class of topological insulating phases was theoretically proposed in 2008, they have attracted significant interest owing to their conceptual novelty, connection to knot theory, and many fascinating physical properties.
However, because their realization requires special forms of long-range spin-orbit coupling (SOC), they have not been achieved in any 3D system yet.
Here we report the first experimental realization of the long-sought-after Hopf insulator in a 3D circuit system.
To implement the Hopf insulator, we construct basic pseudo-spin modules and connection modules that can realize $2\times2$-matrix elements and then design the circuit network according to a tight-binding Hopf insulator Hamiltonian constructed by the Hopf map.
By simulating the band structure of the designed circuit network and calculating the Hopf invariant, we find that the circuit realizes a Hopf insulator with Hopf invariant equaling $4$.
Experimentally, we measure the band structure of a printed circuit board and find the observed properties of the bulk bands and topological surface states (TSS) are in good agreement with the theoretical predictions, verifying the bulk-boundary correspondence of the Hopf insulator.
Our scheme brings the experimental study of Hopf insulators to reality and opens the door to the implementation of more unexplored topological phases beyond the known topological classifications.

\end{abstract}

\maketitle

In 2008, the pioneering tenfold-way classification based on non-spatial symmetries provided the first systematic understanding of non-interacting topological phases of matter~\cite{Schnyder2008,kitaev2009periodic}, and founded the basis for the later discovery of
a long list of symmetry-protected topological phases based on powerful methods such as symmetry indicators~\cite{Chiu2015RMP, Po2017SI, Bradlyn2017, Zhang2019, Tang2019, Vergniory2019}.
Despite its systematicity and fundamental significance, the existence of topological phases beyond the tenfold way classification was soon noticed.
Just in the same year, Moore, Ran, and Wen theoretically showed that a class of 3D two-band magnetic topological insulators~\cite{Moore2008hopf}, later dubbed Hopf insulators as characterized by an integer-valued 
Hopf invariant~\cite{Deng2013hopf, Kennedy2016hopf}, exist outside the tenfold-way periodic table~\cite{Schnyder2008, kitaev2009periodic}.
Besides the prominent conceptual significance,  the two-band Hopf insulators have attracted considerable interest both in theory and experiment due to their many fascinating properties~\cite{Lapierre2021hopf,Alexandradinata2021hopf}.
The bulk-boundary correspondence, a central property of topological phases, is also unique in Hopf insulators.
The uniqueness is manifested through the dependence of TSS on the surface's orientation and the support of gapless surface Dirac cones, even though the time-reversal symmetry is broken.
Besides enriching topological
phases, the study of Hopf insulators also substantially advances the understanding of
2D out-of-equilibrium topological phases. The Hopf invariant is found to
play an important role in the topological characterization of quenched
Chern insulators~\cite{Quench_Chern2017,Chern_linking2019},
quenched Euler insulators~\cite{Euler2020,Simulation_Euler2022},
and Floquet Chern insulators~\cite{Hopf_Floquet2019}.

Although Hopf insulators have been proposed for more than one decade and the great importance of their physical realization is well appreciated~\cite{Moore2008hopf, Schuster2021hopfa, Schuster2021hopfb}, to date they have only been simulated in a single-qubit quantum simulator~\cite{Yuan2017hopf} and have not been implemented in any 3D system yet.
The main challenges for implementing Hopf insulators are the demand of having exactly two bands and a peculiar pattern of long-range SOC.
These requirements rule out the implementation in many quantum material systems as well as many artificial systems.

In this paper, we report the first bulk realization of the long-sought-after Hopf insulators in a 3D circuit.
Because of the extremely high level of connection freedom, circuit networks have been used to realize many novel states of matter, such as 2D topological insulators~\cite{circuit_PRX2015, circuit_PRL2015}, 3D topological semimetals~\cite{circuit_Weyl_CP_2018,circuit_Weyl_CP_2018_yu,circuit_Weyl_prb_2019}, and even 4D topological phases~\cite{circuit4D_yu, PhysRevB4DEzawa, circuit4D_prb,CKT_2020_4D_yidongchong}.
To carry out the experiment, we use basic building blocks, which in principle admit the implementation of any arbitrary two-band model, Hermitian or non-Hermitian, to design a 3D periodic circuit according to a Hopf insulator model constructed by the Hopf map.
By numerically simulating the band structure and calculating the Hopf invariant $N_{h}$, we find a Hopf insulator phase with $N_{h}=4$ exists in a sizable region of the parameter space.
By experimentally measuring the bulk and boundary energy spectra of a printed circuit board sample, we find the experimental results agree well with the theoretical predictions and verify the defining bulk-boundary correspondence of the Hopf insulator.


\textit{Model Hamiltonian.---}We start with the theoretical model for two-band Hopf insulators. It is known that any two-band model can be expressed via the Pauli matrices $\boldsymbol{\sigma}$=$(\sigma_{1}$,$\sigma_{2}$,$\sigma_{3})$ as
\begin{equation}
H(\bm{k})=d_{0}(\bm{k})\sigma_{0}+\bm{d}(\bm{k})\cdot\boldsymbol{\sigma},
\end{equation}
where $\sigma_{0}$ is the $2\times2$ identity matrix and $\bm k=(k_x,k_y,k_z)$.
Focusing on band topology, all essential information is encoded in the three-component $\bm{d}$-vector. The first term on the right-hand side is irrelevant and can be neglected. Moore, Ran, and Wen showed that theoretical models for two-band Hopf insulators can be systematically constructed when the $\bm{d}$ vector is descended from a complex spinor via the Hopf map~\cite{Moore2008hopf}, i.e., $\bm{d}(\bm{k})=z(\bm{k})^{\dagger}\bm{\sigma}z(\bm{k})$, where $z(\boldsymbol{k})=(z_{1}(\boldsymbol{k}),z_{2}(\boldsymbol{k}))^{T}$,
$z_{1}=\eta_{1}(\bm{k})+i\eta_{2}(\bm{k})$, $z_{2}=\eta_{3}(\bm{k})+i\eta_{4}(\bm{k})$,
with $\eta_{1,2,3,4}(\bm{k})$ being real functions of momentum. 
The map is characterized by the Hopf invariant defined as ~\cite{Moore2008hopf}
\begin{equation}
N_{h}=-\frac{1}{4\pi^{2}}\int d^{3}k\epsilon^{\mu\nu\rho}A_{\mu}\partial_{\rho}A_{\nu},\label{eq:-4}
\end{equation}
where $A_{\mu}=-i\langle u(\bm{k})|\partial_{\mu}|u(\bm{k})\rangle$ with $\mu,\nu,\rho=\{k_{x},k_{y},k_{z}\}$ and $|u\rangle$ being the negative-energy eigenfunction of $H(\bm{k})$ is the Berry connection.
It is worth noting that $N_{h}$ does not have any gauge ambiguity even though the Berry connection $A_{\mu}$ is gauge dependent.
Such a property allows us to numerically calculate the Hopf invariant through discretization of the Brillouin zone~\cite{Yan2017hopf}.
When $N_{h}$ is nonzero, the resulting $\bm{d}(\bm{k})\cdot\boldsymbol{\sigma}$ model realizes a two-band Hopf insulator.

There are infinite choices for $\eta_{1,2,3,4}(\bm{k})$ to achieve a nonzero $N_{h}$.
For the convenience of experimental implementation, in this work we consider $\eta_{1}(\bm{k})=t_{1}\sin k_{x}$, $\eta_{2}(\bm{k})=t_{2}\cos(k_{x}+k_{y}+k_{z})$, $\eta_{3}(\bm{k})=t_{3}\sin k_{y}$, and $\eta_{4}(\bm{k})=t_{4}\sin k_{z}$.
Accordingly, we find $N_{h}=4$ and
\begin{eqnarray}
d_{1}(\bm{k})&=&2t_{1}t_{3}\sin k_{x}\sin k_{y}+2t_{2}t_{4}\cos k_{d}\sin k_{z},\nonumber\\
d_{2}(\bm{k})&=&2t_{1}t_{4}\sin k_{x}\sin k_{z}-2t_{2}t_{3}\cos k_{d}\sin k_{y},\nonumber\\
d_{3}(\bm{k})&=&t_{1}^{2}\sin^{2}k_{x}+t_{2}^{2}\cos^{2}k_{d}-t_{3}^{2}\sin^{2}k_{y}-t_{4}^{2}\sin^{2}k_{z}.\quad \label{eq:2x2H}
\end{eqnarray}
Here we have introduced $k_{d}\equiv k_{x}+k_{y}+k_{z}$ to shorten the notation.
Apparently, all three components of the $\bm{d}$ vector involve long-range hopping processes in real space.
What raises a particular challenge is that the hopping parameters involving different length scales need to be comparable in magnitude and satisfy a stringent phase pattern.

		\begin{figure}[t]
		\centering \includegraphics[width=0.996\columnwidth]{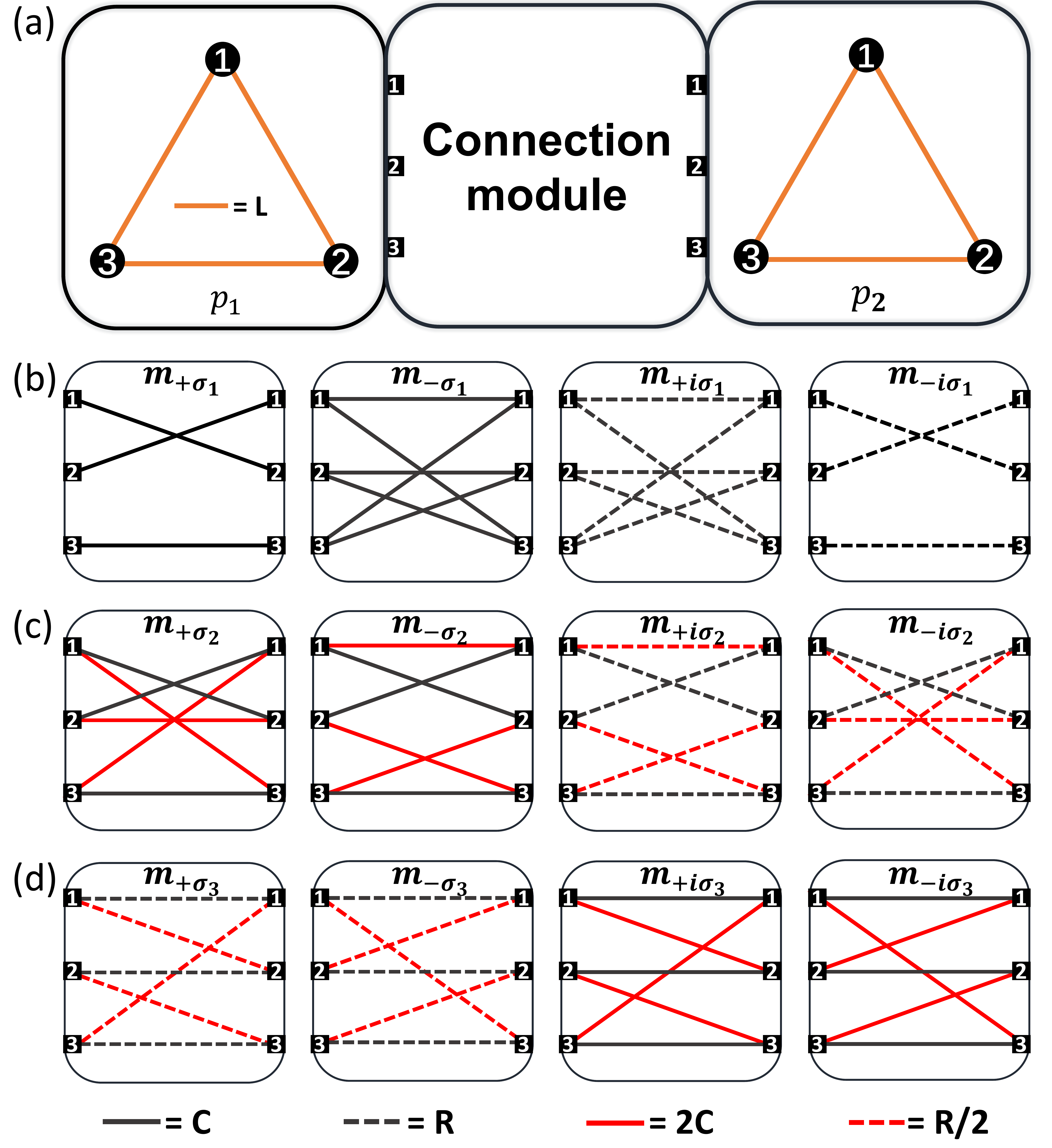}
		\caption{\label{fig:BuildingBlock}
		(a) A connection module connects pseudo-spin modules $p_1$ and $p_2$. The three ports on the connection module's left (right) side are connected to the three nodes of the left (right) pseudo-spin module.
		(b-d) The list of designed connection modules $m_{\pm(i)\sigma_{1,2,3}}$ that gives $\pm(i)\sigma_{1,2,3}$ types of tunneling matrices between the pseudo-spin space, where solid lines indicate capacitors, dashed lines indicate resistors. In each module, the red solid and dashed lines indicate that their impedances are half those of the black solid and dashed lines. }
		\end{figure}

\textit{Hopf insulator circuit.---}In this paper, we overcome the challenge and implement the Hopf insulator Hamiltonian (\ref{eq:2x2H}) in circuit networks as follows.
We first create the pseudo-spin space via the module depicted in Fig.~\ref{fig:BuildingBlock}(a), where three identical inductors form a triangle with ${\rm C}_{3}$ rotational symmetry. Accordingly, the pseudo-spin space in Eq.(\ref{eq:2x2H}) is provided by the twofold degenerate eigenstates characterized by the 2D representation of the ${\rm C}_{3}$ group.
Connecting pseudo-spin modules with connection modules, in which the components form a braided network, allows electrical signals to flip the pseudo-spin, resulting in a SOC-like effect when signals are transmitted between the pseudo-spin modules.
Based on this idea, we design connection modules to generate couplings of the form $\pm(i)\sigma_n$ ($n=1,2,3$)  as shown in Figs.~\ref{fig:BuildingBlock}(b-d)~\cite{NonAbelian_NE_2022,sm}.
As the parameters of the capacitors, inductors, and resistors are positive real numbers, we incorporate the negative sign and the imaginary unit $i$ into the network structure of the connection modules in order to obtain the hopping matrix with complex coefficients in Eq.~(\ref{eq:H_TB}) below.
With these modules, the challenging long-range SOC in Eq.~(\ref{eq:2x2H}) can be achieved since components in electronic circuits can be connected between nodes at arbitrary distances by wires, which is in sharp contrast to condensed solid materials and many artificial materials.
\begin{table}[t]
\begin{centering}
\begin{tabular}{cccc}
\hline
\hline
\textit{$m$} & 0 & 1 & 2\tabularnewline
$\delta_{m}$ & $0$ & $2\hat{x}$ & $2\hat{y}$\tabularnewline
\textit{$\hat{U}_{m}$} & $\frac{1}{2}$$($$t_{1}^{2}$$+$$t_{2}^{2}$$-$$t_{3}^{2}$$-$$t_{4}^{2}$$)$$\sigma$$_{1}$ & $-\frac{1}{4}t{}_{1}^{2}\sigma_{1}$ & $\frac{1}{4}t{}_{3}^{2}\sigma_{1}$\tabularnewline
\hline
\textit{$m$} & 3 & 4 & 5\tabularnewline
$\delta_{m}$ & $2\hat{z}$ & $\hat{x}-\hat{y}$ & $\hat{x}+\hat{y}$\tabularnewline
\textit{$\hat{U}_{m}$} & $\frac{1}{4}t{}_{4}^{2}\sigma_{1}$ & $\frac{1}{2}t_{1}t_{3}\sigma_{3}$ & $\frac{1}{2}(it_{2}t_{4}-t_{1}t_{3})\sigma_{3}$\tabularnewline
\hline
\textit{$m$} & 6 & 7 & 8\tabularnewline
$\delta_{m}$ & $\hat{x}-\hat{z}$ & $\hat{x}+\hat{z}$ & $\hat{x}+2\hat{y}+\hat{z}$\tabularnewline
\textit{$\hat{U}_{m}$} & $\frac{1}{2}t_{1}t_{4}\sigma_{2}$ & $-\frac{1}{2}(it_{2}t_{3}+t_{1}t_{4})\sigma_{2}$ & $\frac{i}{2}t_{2}t_{3}\sigma_{2}$\tabularnewline
\hline
\textit{$m$} & 9 & 10 & \tabularnewline
$\delta_{m}$ & $\hat{x}+\hat{y}+2\hat{z}$ & $2\hat{x}+2\hat{y}+2\hat{z}$ & \tabularnewline
\textit{$\hat{U}_{m}$} & $-\frac{i}{2}t_{2}t_{4}\sigma_{3}$ & $\frac{1}{4}t{}_{2}^{2}\sigma_{1}$ & \tabularnewline
\hline
\hline
\end{tabular}\par\end{centering}
\caption{\label{tab:The-hopping-vectors}The hopping vectors $\delta_{m}$ and the corresponding SOC operators $\hat{U}_{m}$ for the Hopf insulator circuit. $\hat{x}$, $\hat{y}$, and $\hat{z}$ indicate the unit lattice vectors in the x- , y- and z-directions, respectively.}
\end{table}

Performing Fourier transformation on Eq.~(\ref{eq:2x2H}), the real-space tight-binding Hamiltonian reads
\begin{equation}
H_{TB}=\sum_{\bm{l}}\sum_{m=0}^{10}(c_{\bm{l}+\delta_{m},\bm{l}}^{\dagger}\hat{U}_{m}c_{\bm{l}}+h.c.),\label{eq:H_TB}
\end{equation}
where $\bm{l}$ indicate lattice sites, $\delta_{m}$ are hopping vectors, $\hat{U}_{m}$ are SOC operators as given in Table (\ref{tab:The-hopping-vectors}). The operators $\hat{U}_{m}$ can be implemented with the connection modules illustrated in Figs.~\ref{fig:BuildingBlock}(b-d).
For example, $\hat{U}_{1}$ can be built by $m_{-\sigma_{1}}$ module and $\hat{U}_{5}$ can be constructed by connecting the  $m_{-\sigma_{3}}$ and $m_{i\sigma_{3}}$ modules in parallel.
To reduce the number of operational amplifiers used in the experiment, we have exchanged the expressions for $d_{1}$ and $d_{3}$.
This operation does not change the band topology since it is equivalent to a redefinition of the spin basis.

		\begin{figure}[b]
		\centering \includegraphics[width=0.99\columnwidth]{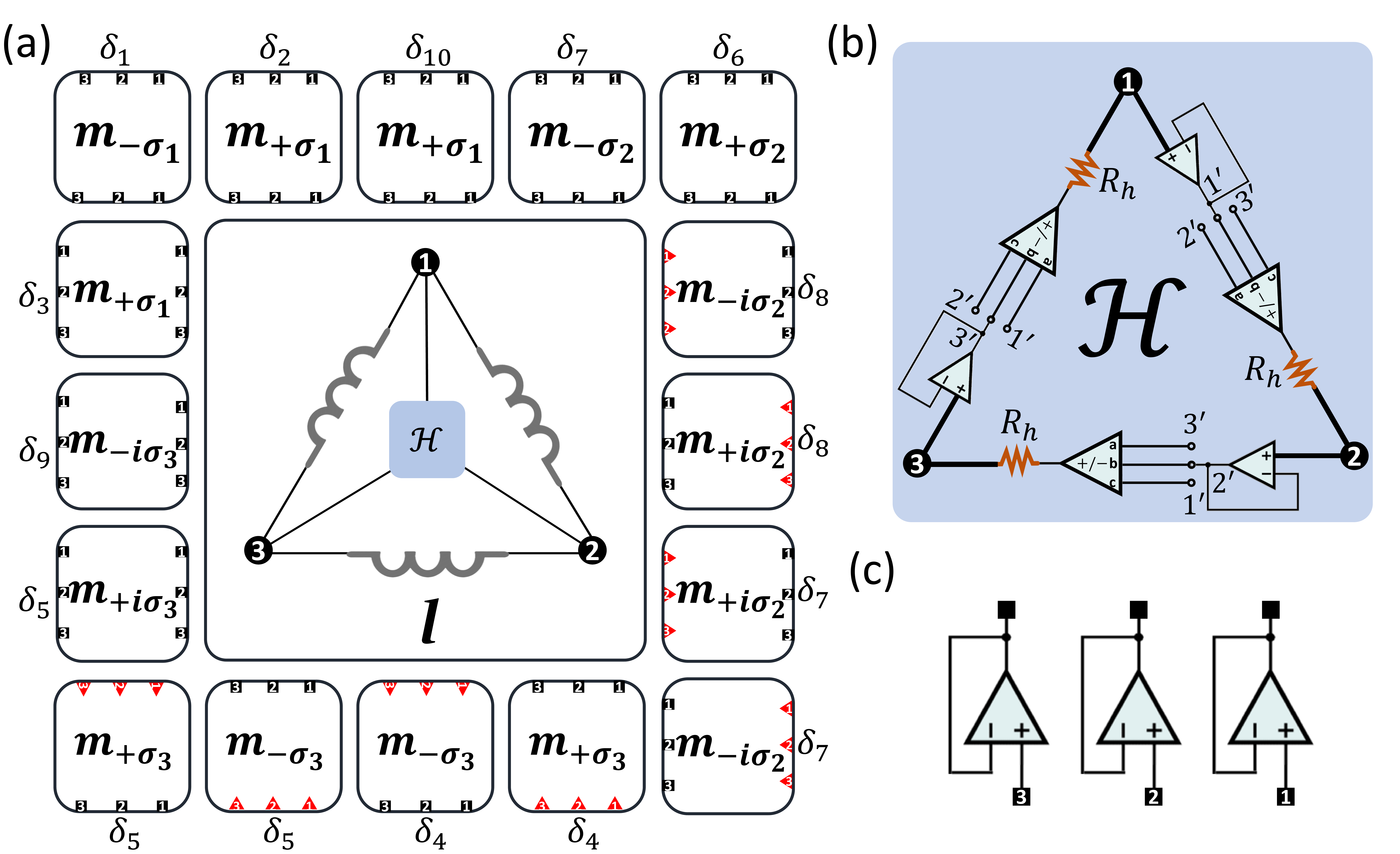}
		\caption{\label{fig:HopfCircuit}
		(a) The unit cell of the Hopf insulator circuit. The connection modules are used for 3D connections. The ports of the connection modules marked with $\delta_m$ ($m=$ 1 to 10) indicate that they are connected to the pseudo-spin module in cell $\bm{l}+\delta_m$, while unmarked ones indicate that they are connected to the pseudo-spin module in cell $\bm{l}$.
		(b) The $\mathcal{H}$-module consists of voltage followers, resistors, and adder-subtractor operational amplifiers.
		(c) Detailed circuit diagram of the red ports in Fig.~\ref{fig:HopfCircuit}(a).}
		\end{figure}

According to Eq.~(\ref{eq:H_TB}), we construct the 3D Hopf insulator circuit network using pseudo-spin modules and connection modules.
The unit cell of the Hopf insulator circuit is shown in Figs.~\ref{fig:HopfCircuit}(a-c). 
It is worth noting that the resistors in the modules $m_{\pm i\sigma_{2}}$ and $m_{\pm\sigma_{3}}$ induce energy loss, which makes the Hamiltonian non-Hermitian.
To address this issue, we use the $\mathcal{H}$-module (Fig.~\ref{fig:HopfCircuit}(b)) to compensate for the energy loss and restore the hermiticity~\cite{sm}.

Kirchhoff's equations for the Hopf insulator circuit can be written as
\begin{equation}
(h_{1}(\bm{k})\oplus H_{h}^{circuit}(\bm{k}))\tilde{\bm v}=\omega^{-2}(0\oplus I_{2})\tilde{ \bm v}\label{eq:-x3}
\end{equation}
where $\oplus$ stands for a direct sum of the constant representation space and the pseudo-spin space of the ${\rm C}_{3}$ symmetry group,  $\tilde{\bm v}=U^{\dagger}{\bm v}$, $\bm v=({v}_{1},{v}_{2},{v}_{3})^{T}$ are the node voltages in the unit cell, and $U$ is defined in Supplemental Material~\cite{sm}.
$h_{1}(\bm{k})$ is the Hamiltonian in the constant representation space.
$H_{h}^{circuit}(\bm{k})=\sum_{i=0}^{3}f_{i}(\bm{k})\sigma_{i}$ is the Hamiltonian in the pseudo-spin space, where
\begin{eqnarray}
f_{0}(\bm{k})&=&\frac{L}{3}\big(4C_{1}+2C_{2}+2C_{3}+2C_{4}+12(C_{5}+C_{6})\big),\nonumber\\
f_{1}(\bm{k})&=&-\frac{4L}{3}\big(C_{1}\sin^{2}k_{x}+C_{2}\cos^{2}k_{d}
-C_{3}\sin^{2}k_{y}\nonumber\\
&&-C_{4}\sin^{2}k_{z}\big),\nonumber\\
f_{2}(\bm{k})&=&-\frac{4\sqrt{3}L}{3}\big(-\frac{1}{R_{1}\omega}\cos k_{d}\sin k_{y}
+C_{5}\sin k_{x}\sin k_{z}\big),\nonumber\\
f_{3}(\bm{k})&=&-\frac{4\sqrt{3}L}{3}\big(C_{6}\cos k_{d}\sin k_{z}
+\frac{1}{R_{2}\omega}\sin k_{x}\sin k_{y}\big),
\label{eq:-x4}
\end{eqnarray}
and $R_{\alpha}$ ($\alpha=1,2$), $C_{\beta}$ ($\beta=1$ to $6$), $L$ are parameters of the components in the circuit.
By choosing appropriate parameters, one can separate the eigenfrequencies of $H_{h}^{circuit}(\bm{k})$ well from that of $h_{1}(\bm{k})$.
Therefore, we focus on $H_{h}^{circuit}(\bm{k})$ and examine its topological properties below.

        \begin{figure}[t]
		\centering \includegraphics[width=0.99\columnwidth]{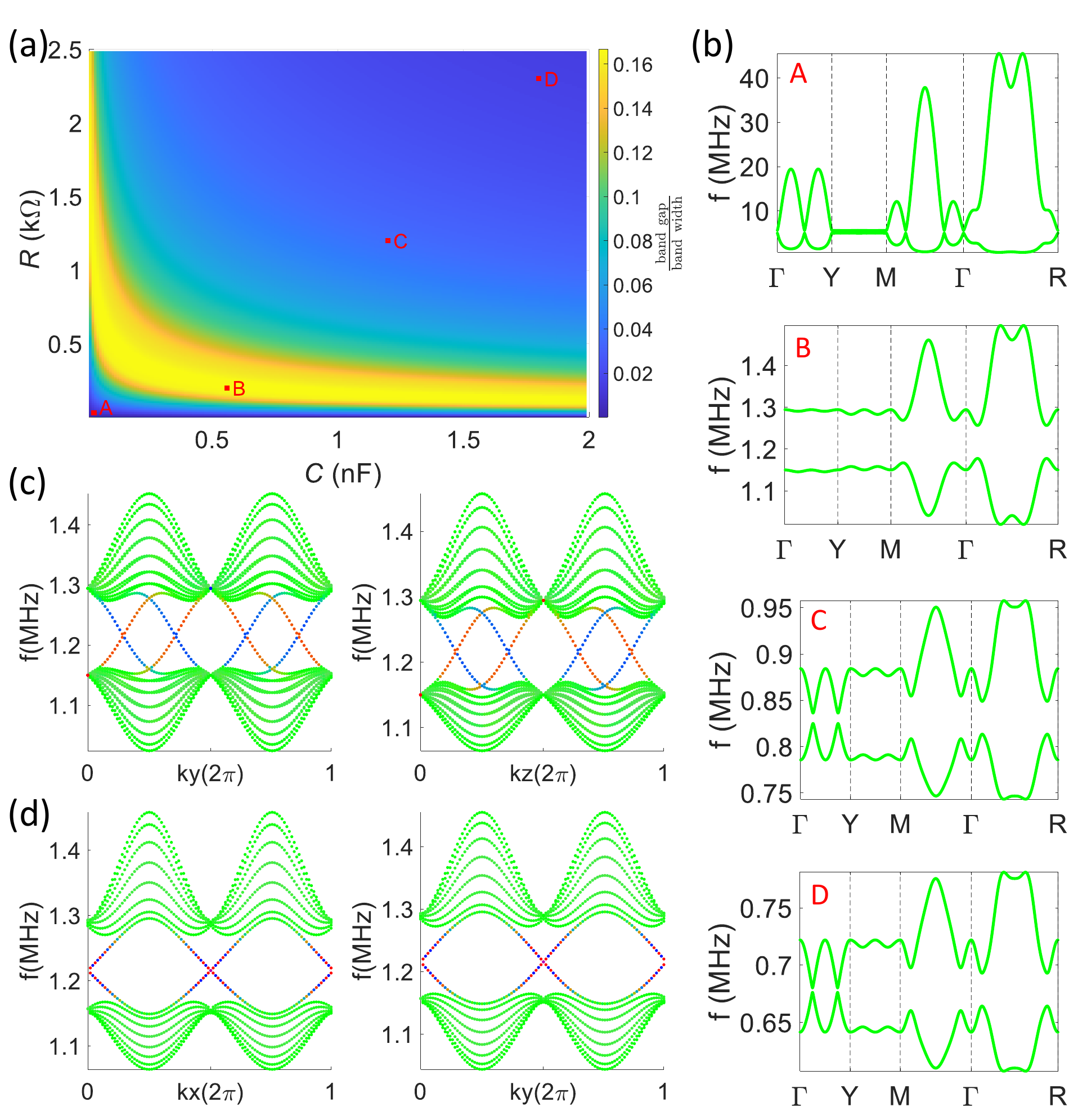}
		\caption{\label{fig:ModelResults}
		(a) The calculated band gap as a function of the parameters of  $C$ and  $R$. The color map represents the ratio of the band gap to the total frequency band width.
		(b) The frequency spectrum of $H_{h}^{circuit}(\bm{k})$ at points A, B, C, and D in the phase diagram, where $\Gamma$=(0,0,0), $\rm Y$=(0,1,0), $\rm M$=(1,1,0), $\rm R$=(1,1,1) are the high symmetry points in the Brillouin zone in units of $\pi$ with the unit cell lattice parameters set to 1. The parameters ($R$, $C$) are equal to (0.03 $\rm k\Omega$, 0.03 nF) at A point, (0.2 $\rm k\Omega$, 0.56 nF) at B point, (1.2 $\rm k\Omega$, 1.2 nF) at C point, and (2.3 $\rm k\Omega$, 1.8 nF) at D point. The inductance is fixed as $L$=2.7 ${\rm \mu}$H in all calculations.
		(c) The calculated frequency spectrum along $k_{y}$ and $k_{z}$ directions of a 16-layers slab with x-direction surfaces. The green color refers to bulk states. The red (blue) color refers to states localized on the x=1 (x=16) layer.
		(d) The frequency spectrum of a 16-layers slab with z-direction surfaces. The TSS on the z=1 (red) and z=16 (blue) layers degenerate and overlap in frequency.
		The parameters at point B are used to calculate the TSS in Figs.~\ref{fig:ModelResults}(c-d).}
		\end{figure}

For simplicity of discussion, we set the inductance $L$=2.7 $\mu$H,  $R_\alpha$=$R$ ($\alpha=1,2$),  and $C_\beta$=$C$ ($\beta=1$ to $6$).
The phase diagram of the frequency band gap as a function of $R$ and $C$ is shown in Fig.~\ref{fig:ModelResults}(a), where the band gaps are finite in the yellow and green regions and tend to zero in the dark blue region as shown in Fig.~\ref{fig:ModelResults}(b).
In the gapped region, we find $N_{h}=4$ as expected, agreeing with the considered Hopf map.
In the regions with vanishingly small gaps, the Hopf invariant does not converge as it is not well-defined in the presence of band degeneracy.

One defining characteristic of the Hopf insulator is the unique correspondence between the number of topologically protected TSS and the Hopf invariant.
Measuring this bulk-surface correspondence can faithfully identify the topological nature of the system.
In Figs.~\ref{fig:ModelResults}(c-d), we show the numerically-calculated band structures for a 16-layers thickness slab structure terminated in x- and z-direction, respectively.
There are four TSS on the x-direction surface of the system in Fig.~\ref{fig:ModelResults}(c), where the red (blue) color refers to TSS located on the x=1 (x=16) layer, and the green color refers to bulk states.
On the z-direction termination, the TSS localized on z=1 and z=16 layers are degenerate in frequency, as shown in Fig.~\ref{fig:ModelResults}(d).
Remarkably, on each z-direction surface, the crossings of the surface frequency spectra labeled by the same color at time-reversal invariant momenta reveal the existence of surface Dirac cones even though the Hamiltonian does not have time-reversal symmetry as already mentioned.
Experimentally, these exotic surface Dirac cones provide a unique signature to identify the Hopf insulator.

In the following, we experimentally verify the topological nature of the Hopf insulator designed above.
According to the circuit diagram in Figs.~\ref{fig:HopfCircuit}(a-c), we prepare a printed circuit board with the number of unit cells being 3$\times$20$\times$6 in x-, y-, and z-directions and set periodic boundary conditions in the x- and y-directions and open boundary conditions in the z-direction. The circuit structure of the unit cell and the $\mathcal{H}$-module are shown in Figs.~\ref{fig:ExpResults}(a-b). A global view of the printed circuit board is shown in Supplemental Material~\cite{sm}.

		\begin{figure}[b]
		\centering \includegraphics[width=0.99\columnwidth]{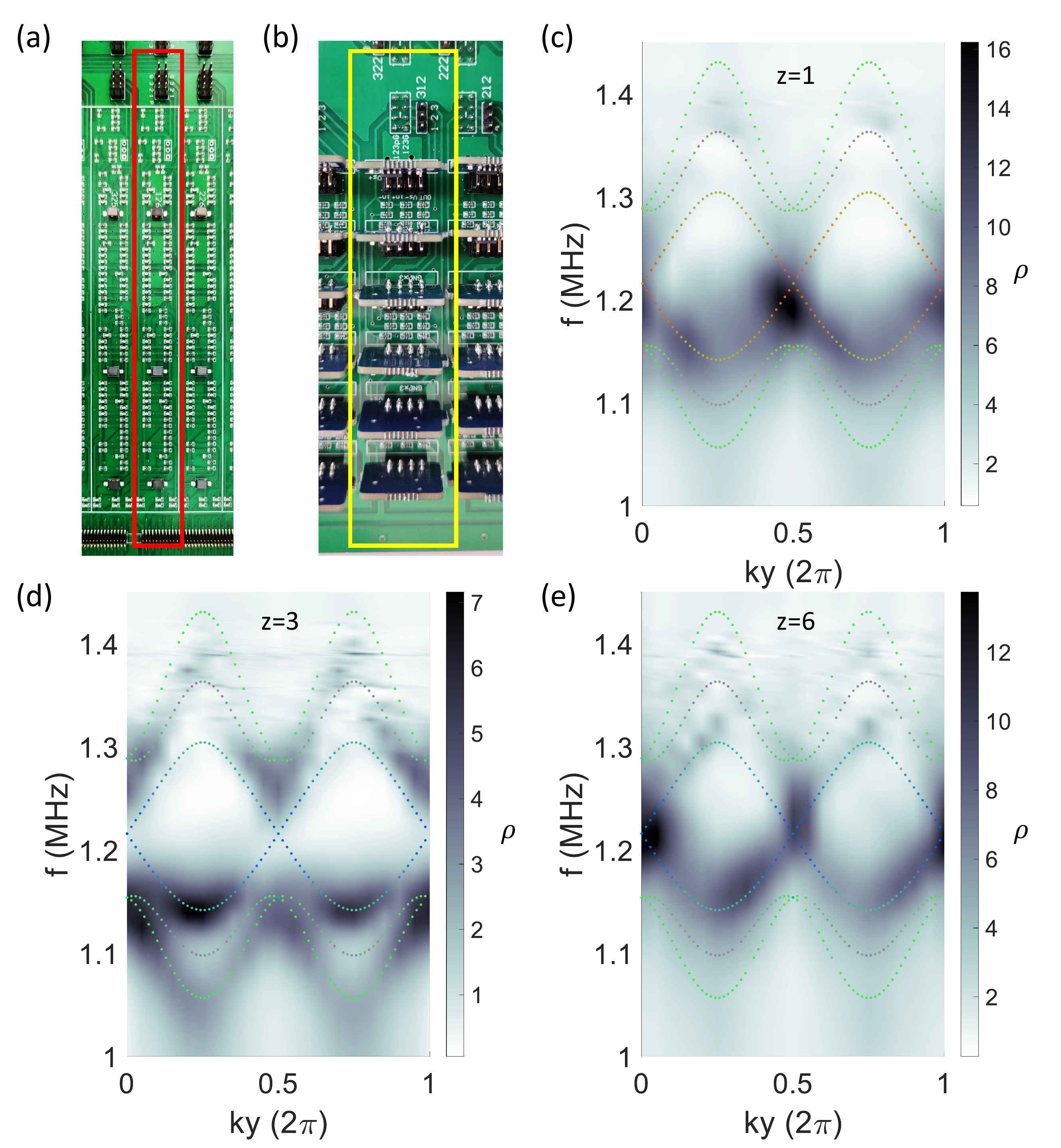}
		\caption{\label{fig:ExpResults}
		(a) The unit cell of the printed circuit board (in the red box) is fabricated according to Fig.~\ref{fig:HopfCircuit}(a).
		(b) The unit cell of the printed circuit board of the $\mathcal{H}$-module (in the yellow box).
		(c), (d) and (e) show the experimentally measured band structure along $k_{y}$ direction with the signal source connected to a cell at z=1 (bottom surface), z=3 (bulk), and z=6 (top surface), respectively. The black color represents the experimental data, and the colored dotted lines are the computed band structure of a slab with six layers in the z-direction,  where red (blue) indicates the wave functions localized on the z=1 (z=6) surface and green indicates the bulk states.}
		\end{figure}

To extract the frequency spectrum of the circuit lattice, we perform frequency-domain measurements to obtain the voltage vectors $\bm v(\bm{r},f)$=$(v_1(\bm{r},f)$, $v_2(\bm{r},f)$, $v_3(\bm{r},f))$, where $\bm{r}$=$(\rm{x},\rm{y},\rm{z})$ labels the unit cell, $f$ is the frequency, and the subscripts indicate the nodes in each unit cell.
By performing Fourier transformation in the x- and y-directions, the voltage $\bm v(\boldsymbol{k}_{\parallel},{\rm z},f)$ can be obtained in the momentum space, where $\boldsymbol{k}_{\parallel}$=$(k_{x},k_{y})$.
The frequency dispersions shown in Figs.~\ref{fig:ExpResults}(c-e)  are obtained by plotting $\rho(\boldsymbol{k}_{\parallel},{\rm z},f)$=$|\bm v(\boldsymbol{k}_{\parallel},{\rm z},f)|^{2}$, where the peaks of  $\rho$ indicate the resonance frequency of the circuit system.

The frequency dispersions for the excitation signal applied to the z=1 and z=6 layers are depicted in Fig.~\ref{fig:ExpResults}(c) and (e), where the TSS appear in the band gap and locate around the time-reversal invariant momenta, which is in good agreement with the results calculated from the model Hamiltonian (color dotted lines).
The TSS disappear for the excitation signal applied to the z=3 layer because the signal on the middle layer cannot excite the TSS located on the surface layers, as shown in Fig.~\ref{fig:ExpResults}(d).
These experimental results confirm the predicted properties of the TSS.
In Figs.~\ref{fig:ExpResults}(c-e), we have shifted the experimental data upward by 2.9$\times10^{4}$ Hz to compare with the theoretical results.
Details of the distribution of the TSS in the z-direction are provided in the Supplemental Material~\cite{sm}.
As a final remark, the TSS in the circuit lattice can also be detected by other methods, like impedance measurements~\cite{HOTAI2021,hyperbolic2022}.

\textit{Conclusions and discussions.---}In this work, we present a general scheme for the implementation of long-range SOC in electric circuits, where the spatial dependence of the SOC can be modulated with a high degree of freedom.
Using this property, we have successfully implemented the long-sought-after Hopf insulator in the circuit and observed the TSS enforced by bulk-boundary correspondence.
Our general scheme brings the experimental study of Hopf insulators to reality and paves the way for exploring other exotic topological phases associated with peculiar SOC.
With our established platforms, the experimental exploration of links and knots with very rich topological structures becomes accessible~\cite{Yan2017link, Bi2017knot, Belopolski2022}.
Moreover, the idea behind our scheme can be applied to the future design of systems with more but fixed bands to implement the novel topological phases beyond the known topological classifications, such as Hopf insulators in three-band systems~\cite{Neupert2012} and models constructed by higher-dimensional generalizations of the Hopf map~\cite{Liu2017hopf}.

\textit{Acknowledgments.---}R. Y. was supported by the National Key Research and Development Program of China (No.2017YFA0304700, No.2017YFA0303402), the National Natural Science Foundation of China (No.11874048 and No.12274328), and the Beijing National Laboratory for Condensed Matter Physics. Z. Y. was supported by the National Natural Science Foundation of China (Grant No.11904417 and No.12174455) and the Natural Science Foundation of Guangdong Province (Grant No. 2021B1515020026).

\bibliography{refs_yan}

\end{document}